\documentclass[
 aip, 
reprint,
]{revtex4-1}

\usepackage[utf8]{inputenc}
\usepackage[T1]{fontenc}
\usepackage{diagbox}
\usepackage{amsfonts}
\usepackage{amsmath}
\usepackage{amsthm}
\usepackage{xr}
\usepackage{bm}
\usepackage{indentfirst}
\usepackage{microtype}
\usepackage[squaren]{SIunits}
\usepackage{booktabs}
\usepackage{amssymb}
\usepackage{graphicx}
\usepackage{multirow}
\usepackage{wallpaper}
\usepackage[normalem]{ulem}

\newcommand{\beq}{\begin{eqnarray}}
\newcommand{\eeq}{\end{eqnarray}}
\newcommand{\bea}{\begin{eqnarray}}
\newcommand{\eea}{\end{eqnarray}}

\newcommand \pvec{{\bf p}}

\newcommand\Rvec{{\bf R}}

\usepackage{braket}
\usepackage{caption}
\usepackage{mathtools}
\usepackage{subcaption}
\usepackage[colorlinks=true,urlcolor=blue,linkcolor=blue,
            citecolor=blue]{hyperref}
\usepackage[normalem]{ulem}
\newcommand\redsout{\bgroup\markoverwith{\textcolor{red}{\rule[.5ex]{2pt}{0.4pt}}}\ULon}

\usepackage{xr}
\makeatletter
\newcommand*{\addFileDependency}[1]{
  \typeout{(#1)}
  \@addtofilelist{#1}
  \IfFileExists{#1}{}{\typeout{No file #1.}}
}
\makeatother

\newcommand*{\myexternaldocument}[1]{
    \externaldocument{#1}
    \addFileDependency{#1.tex}
    \addFileDependency{#1.aux}
}

\myexternaldocument{SupportingInformation}

\def\simge{\mathrel{%
       \rlap{\raise 0.511ex \hbox{$>$}}{\lower 0.511ex \hbox{$\sim$}}}}
\def\simle{\mathrel{
       \rlap{\raise 0.511ex \hbox{$<$}}{\lower 0.511ex \hbox{$\sim$}}}}

\begin{document}


\title{Multi-scale simulation of the adsorption of lithium ion on graphite surface: from Quantum Monte Carlo to Molecular Density Functional Theory }

\author{Michele Ruggeri}
\affiliation{Maison de la Simulation, CEA, CNRS, Univ. Paris-Sud, UVSQ, Universit{\'e} Paris-Saclay, 91191 Gif-sur-Yvette, France}
\author{Kyle Reeves}
\affiliation{Maison de la Simulation, CEA, CNRS, Univ. Paris-Sud, UVSQ, Universit{\'e} Paris-Saclay, 91191 Gif-sur-Yvette, France}
\author{Tzu-Yao Hsu}
\affiliation{ 
Sorbonne Universit\'e, CNRS, Physicochimie des \'Electrolytes et Nanosyst\`emes Interfaciaux, F-75005 Paris, France
}%
\author{Guillaume Jeanmairet}
\affiliation{ 
Sorbonne Universit\'e, CNRS, Physicochimie des \'Electrolytes et Nanosyst\`emes Interfaciaux, F-75005 Paris, France
}%
\author{Mathieu Salanne}
\email{mathieu.salanne@sorbonne-universite.fr}
\affiliation{Maison de la Simulation, CEA, CNRS, Univ. Paris-Sud, UVSQ, Universit{\'e} Paris-Saclay, 91191 Gif-sur-Yvette, France}
\affiliation{ 
Sorbonne Universit\'e, CNRS, Physicochimie des \'Electrolytes et Nanosyst\`emes Interfaciaux, F-75005 Paris, France
}%
\affiliation{Institut Universitaire de France (IUF), 75231 Paris, France}
\author{Carlo Pierleoni}
\email{carlo.pierleoni@aquila.infn.it}
\affiliation{Maison de la Simulation, CEA, CNRS, Univ. Paris-Sud, UVSQ, Universit{\'e} Paris-Saclay, 91191 Gif-sur-Yvette, France}
\affiliation{Department of Physical and Chemical Sciences, University of L'Aquila, Via Vetoio 10, I-67010 L'Aquila, Italy}
\date{\today}

\begin{abstract}
The structure of the double-layer formed at the surface of carbon electrodes is governed by the interactions between the electrode and the electrolyte species. However, carbon is notoriously difficult to simulate accurately, even with well-established methods such as electronic Density Functional Theory  and Molecular Dynamics. Here we focus on the important case of a lithium ion in contact with the surface of graphite, and we perform a series of reference Quantum Monte Carlo  calculations that allow us to benchmark various electronic Density Functional Theory functionals. We then fit an accurate carbon--lithium pair potential, which is used in molecular Density Functional Theory  calculations to determine the free energy of the adsorption of the ion on the surface in the presence of water. The adsorption profile in solution differs markedly from the gas phase results, which emphasize the role of the solvent on the properties of the double-layer.
\end{abstract}

\maketitle

\section{\label{sec:Introduction}Introduction}

Carbon materials play a very important role in many chemistry fields, ranging from catalysis to electrochemistry. In energy storage applications, carbon is used as an electrode in the form of graphite for Li-ion batteries,~\cite{armand2008a} of hard carbon for Na-ion batteries,~\cite{dou2019a} and of nanoporous materials for supercapacitors.~\cite{simon2012a} In all these examples, the carbon materials are put in contact with a charged electrolyte, and the interfacial structure and dynamics play a crucial role in the operation of the devices.~\cite{salanne2016a,striolo2016a} In order to optimize the performance, molecular simulations can play an important role,~\cite{lahrar2021a}  but they need to accurately account for the interactions between the ions and the carbon surface.

Over the years, a large number of electronic density functional theory (eDFT)  and molecular dynamics (MD) simulations were devoted to the study of electrochemical interfaces between carbon and ions or liquid electrolytes. Although they provide qualitatively similar results, some discrepancies may be observed when analyzing quantitative properties, such as adsorption energies, preferential ``binding'' distances or adsorption profiles. For example, as reported by Valencia {\it et al.}\cite{valencia2006a}, in the case of bare lithium adsorption on graphite surface, eDFT calculations always display binding energies larger than 1~eV with a preference for the center of the C$_6$ hexagonal rings (hollow sites) with respect to adsorption above C atoms (top sites) or C--C bonds (bridge sites).~\cite{valencia2006a} The lithium atom loses a valence electron which is entirely transferred to the carbon surface. Yet, the adsorption energy may differ by more than 0.5~eV when changing the exchange-correlation functional.~\cite{rytkonen2007a} In particular, the inclusion of dispersion effect further increases the binding energy of the lithium ion.~\cite{fan2013a}
 
Similar problems arise in classical MD simulations due to the choice of different interaction potentials. On the one hand, the choice of the electrostatic model for the electrode atoms, which can be treated using fixed-charges or the constant potential method,~\cite{scalfi2021a} will impact the localization of the charge induced by the ions on the carbon. On the other hand, several force fields may be used for the short-range repulsion and dispersion interactions.

Important discrepancies were already reported in the case of the adsorption of water molecules on carbon surfaces.~\cite{striolo2016a} 
In two studies on carbon nanotubes~\cite{alhamdani2017a} and other carbon nanostructures~\cite{brandenburg2019a}, Michaelides and co-workers compared the results  of eDFT calculations using a large set of exchange-correlation functionals with diffusion Monte Carlo (DMC) calculations. DMC is a Quantum Monte Carlo (QMC) method, which explicitly accounts for electronic correlation and exchange. Although for fermions the fixed node error prevent DMC from being exact, it is more accurate than eDFT and belongs to the class of variational methods that can be systematically improved with enough effort. This method is much more costly than conventional eDFT in terms of computational time, but it is expected to capture with a high accuracy the intermolecular interactions. It can thus be viewed as a good reference for benchmarking purposes. Based on these results\cite{alhamdani2017a,brandenburg2019a}, it appears that the choice of the exchange-correlation (XC) approximation in eDFT is crucial but also that it is difficult to predict the accuracy since the adsorption energy results from a subtle balance of the interactions, especially at medium range.

In this work we follow a similar approach for the adsorption of lithium on graphite. The choice of the system was made based on its relevance for energy applications and because the lithium ions has a small number of valence electrons, thus allowing the cost of the reference DMC calculations. We computed the adsorption profiles on hollow, top and bridge sites by varying the distance between the lithium and the carbon material. DMC results are then compared with several XC approximations. 

Yet, knowing the adsorption energy of a single ion may not be enough. Indeed, the solvent molecules within the electrolyte will also interact strongly both with the surface and with the ions, which can result in large variations of the adsorption properties under realistic conditions.~\cite{merlet2013a} The main quantity to be determined is then the free energy of adsorption profile, which should be obtained by sampling the whole liquid degrees of freedom. This quantity is not accessible to electronic structure methods due to their large computational cost, and it is necessary to resort to classical mechanics-based methods instead. Classical MD is usually the method of choice for such purposes, but it suffers from high inefficiency for systems studied under large dilution conditions. In such cases, it may conveniently be replaced by molecular DFT (MDFT). MDFT describes a liquid by its density field, and a functional of this density is introduced to account for entropic, external and solvent-solvent interactions. The external contribution arises from the species in contact with the liquid, which are the graphite electrode and the lithium ion in the present study. The equilibrium density and its related free energy are computed through a variational procedure, which reduces the computational cost by several orders of magnitude w.r.t atomistic MD. Recently, we have extended the MDFT method to account for constant potential electrodes, and we showed it was providing the correct structure and thermodynamics for liquid water in contact with graphite electrodes.~\cite{jeanmairet_study_2019}

In the second step of the present work, we build upon these developments to study the adsorption of lithium on graphite, in the presence of liquid water. The lithium--carbon interaction is parameterized using the electronic DFT calculations benchmarked on DMC. We show that the energy minimum observed in vacuum vanishes in the presence of solvent, and that the lithium ion does not show anymore a preferred binding distance close to the surface due to the presence of water molecules. 

\section{\label{sec:Methods}Computational methods}
\subsection{Quantum Monte Carlo}
QMC simulations are a set of stochastic computational methods for the evaluation of observables of quantum systems. The fundamental idea behind these techniques is 
that expectation values of physical quantities of quantum systems can be written as
\begin{equation}
\langle A \rangle = \int d \Rvec \, \Psi^*(\Rvec) \hat{A} \Psi(\Rvec) 
\label{eq:qmc_integral}
\end{equation}
where $\hat{A}$ is a generic observable, $\Psi(\Rvec)$ is the (normalized) wave function for the quantum system under analysis , and $\Rvec$ represents the $3N$-dimensional electronic coordinate. The integral in Equation \ref{eq:qmc_integral} can then be evaluated using a Monte Carlo sampling \cite{metropolis1953}.
A Markov chain Monte Carlo simulation consists in sampling a collection of configurations of the system used to estimate configurational integrals such as the one in Equation \ref{eq:qmc_integral}. 

In order to obtain meaningful information on the physical system it is important to use a proper wave function. Since the exact ground state for physical systems of interest is unknown it is necessary to   guess the form of the ground state wave functions. A large variety of QMC methods exists, implementing different strategies. In this study two QMC techniques were used: Variational Monte Carlo (VMC) and DMC, using the fixed-node approximation.

In VMC we approximate the ground state of the system by defining a trial wave function $\Psi_T(\pvec ; \Rvec$), which is then plugged in Equation \ref{eq:qmc_integral}; the resulting integral is then computed using a generalised Metropolis sampling. The trial wave function has a functional form that is chosen by taking into account the physical features of the system, and
depends on a set of variational parameters ${\pvec}$, which are optimized to obtain an approximation of the exact ground state. To optimize the trial wave function we use the variational principle in quantum mechanics: for any given Hamiltonian the ground state has by definition the lowest possible energy. What is done in practice is computing the energy using
the trial wave function
\begin{equation}
\begin{split}
E_{\pvec} &= \int d \Rvec \, \Psi_T^*(\pvec; \Rvec) \hat{H} \Psi_T(\pvec; \Rvec) = \\ &= \int d \Rvec \, \vert\Psi_T(\pvec; \Rvec)\vert^2 E_{loc}(\pvec; \Rvec)    
\end{split}
\label{eq:vmc_energy}
\end{equation}
where we used the local energy $E_{loc}(\pvec;\Rvec)$ defined as
\begin{equation}
E_{loc}(\pvec;\Rvec) = \frac{\hat{H}\Psi_T(\pvec;\Rvec)}{\Psi_T(\pvec;\Rvec)};
\label{eq:eloc}
\end{equation}
By varying the parameters $\pvec$ we can minimize the energy $E_{\pvec}$, thus finding an approximation of the exact ground state within the chosen class the trial functions. Another property that is used in wave function optimization is the zero-variance principle: for a given Hamiltonian, the energy variance of each energy eigenstate is zero. In order to use both the variational and the zero-variance principles the quantity that is optimized in practice is a linear combination of the energy and its variance. 

While VMC allows estimates of ground state properties starting from a relatively simple trial wave function, VMC results are typically not realistically accurate. In order to improve these results projection methods have to be used. Projection methods are a class of QMC methods that use imaginary time evolution to filter out excited states contribution from the trial wave function, allowing accurate computation of ground state properties. While in principle projection techniques are exact, the fermionic sign problem in practice prevents the exact evaluation of quantum observables; to avoid the sign problem the fixed-node approximation is used. With this approximation the results will still be approximate, but they will remain variational, and significantly more accurate than the ones obtained with VMC (i.e. the computed energies and energy variance will be lower).

Several different projection QMC methods exist. Among those, one of the most commonly used in electronic structure calculations is DMC. If we consider a wave function $\Psi(\Rvec)$, a Hamiltonian with a potential energy term $U(\Rvec)$ and we define the imaginary time as $\tau = i t$ in atomic unit we have
\begin{equation}
\hat{H}\Psi(\Rvec) = -\frac{1}{2} \Delta \Psi(\Rvec) + U(\Rvec) \Psi(\Rvec) = \frac{\partial \Psi(\Rvec)}{\partial \tau} ;
\label{eq:imaginary_time}
\end{equation}
the basic idea at the foundation of DMC is that there is a strong analogy between the imaginary time Schr\"odinger equation and a diffusion equation (coming from the kinetic part of $\hat{H}$) with an additional branching term (from $U(\Rvec)$). It is thus possible to simulate the imaginary time propagation by using the process of diffusion and branching of classical random walkers, initially distributed according to a trial wave function (obtained via the optimization procedure described above). The practical details of the implementation of DMC can be found in literature \cite{foulkes2001,umrigar2001}. The important things to stress here is that accuracy of fixed-node DMC result depends on the accuracy of the nodal surface of the trial wave function, and that the results are variational with respect to the nodal location.

All QMC computations in the present work were performed using the QMCpack software \cite{kim2018}.
We used QMC calculations to determine the adsorption energy of Li atom adsorbed on a graphite substrate as a function of the separation between the Li atom from the surface.
If $z$ is the distance between the Li atom and the carbon surface the adsorption energy profile $E_{ads}(z)$ is defined as 
\begin{equation}
E_{ads}(z) = E_{Li+C}(z) - (E_{Li} + E_{C})
\label{eq:bindingenergy}
\end{equation}
where  $E_{Li+C}(z)$ is the energy of a system made of a graphite substrate with a Li atom at a distance $z$ and $E_{Li}$ and  $E_{C}$ are the energy of the isolated atom and graphite respectively. The latter is modeled using two graphene layers made of 50 C atoms each, with an AB stacking, at a distance of 3.47 \AA. 
The energy profile was computed for three different setups: with the Li atom lying above the centre of a C hexagon (hollow site), above a C atom (top site) and above a C--C bond (bridge site).

All simulations were done using a simulation box with cell parameters (in \r{A})
\[\begin{bmatrix}
12.336 &  0.000   &  0.000 \\
6.168  & 10.683   &  0.000 \\
0.000  &  0.000   & 30.000
\end{bmatrix}\]
The height of the simulation box was selected after systematically testing the convergence of the total energy as a function of the amount of vacuum between periodic images in the $z$ direction of the graphite bi-layer using eDFT (Supplementary Figure S1). 
In order to reduce size effects a grid of $4 \times 4 \times 1$ twists was used (corresponding to eight nonequivalent twists) \cite{lin2001}. The same simulation cell and twist grid were used in all the simulations, including the ones of the isolated atom and substrate.

In all VMC and DMC simulations we used trial wave function with a Slater--Jastrow form
\begin{equation}
\Psi_T (\pvec; \Rvec) = J(\pvec; \Rvec)  D(\Rvec)
\end{equation}
where $J(\pvec; \Rvec)$ is a Jastrow term describing electronic correlation, with one and two body terms, and $D(\Rvec)$ is a Slater determinant, ensuring the correct fermionic antisymmetry. The single particle orbitals used in the Slater determinant were evaluated using DFT with a PBE\cite{perdew1996} functional. The orbital calculations were performed using the {\sc Quantum ESPRESSO} software \cite{giannozzi2009,giannozzi2017}. In the DFT calculations a plane wave basis set was used, with a cutoff at 150 Ry, using norm conserving pseudopotentials for both the Li and C atoms. In QMC simulations the Burkatzki--Dolg--Filippi set of pseudopotentials was used \cite{burkatzki2007}.
Wave function optimization is performed using the Linear method \cite{toulouse2007}, and iterated until the optimized energy converges. Only the Jastrow part of the trial wave function is optimized. More information on the Jastrow part of the trial wave function can be found in section S2 of the Supporting Information.
We report DMC results obtained using an imaginary time step of $\tau = 0.01$ Ha$^{-1}$ and a population of 12000 to 16000 random walkers. These choices are consistent with previous works on similar systems \cite{ganesh2014}. Size consistent, non local T--moves\cite{casula2010} are used, as well as the ZSGMA branching scheme\cite{zen20016}.




\subsection{Electronic Density functional theory}

Electronic DFT calculations were performed using the {\sc  Quantum ESPRESSO} electronic structure code.\cite{giannozzi2009,giannozzi2017} To be consistent with the QMC result, an identical simulation cell was considered, consisting of one hundred carbon atoms divided amongst two graphite layers with AB stacking. A kinetic energy cutoff of 40~Ry was used.

We compared the adsorption energy profiles for two series of XC functionals. In the first series we used the LDA,~\cite{ceperley1980a} PBE~\cite{perdew1996} and BLYP~\cite{becke1988a,lee1988a} which neglect the London dispersion interaction. Then we included the latter using either the D2 correction parameterized by Grimme,~\cite{grimme2006a} or through the use of the VDW-DF-C09 functional~\cite{thonhauser2015a,thonhauser2007a,berland2015a} implemented in the Libxc library.~\cite{lehtola2018a} Rappe-Rabe-Kaxiras-Joannopoulos ultrasoft (rrkjus) pseudopotentials~\cite{rappe1990a} were used for both carbon and lithium atoms.

Several uniform Monkhorst-Pack grids of 1$\times$1$\times$1, 2$\times$2$\times$1, 3$\times$3$\times$1 and 5$\times$5$\times$1 $k$-points were tested for a single Li distance of 2.4~\AA\ from the graphite surface (see Supporting Information Section S3). The difference between the total energy of the 1$\times$1$\times$1 and 2$\times$2$\times$1 grids is roughly 26~meV and thus the 1$\times$1$\times$1 grid was used for the calculations throughout this work. Additionally, an extended system consisting in four carbon layers instead of two was simulated to check the effect of the number of graphite layers; almost no difference was observed for the free energy profile as shown in Supplementary Section 4. 

Like for the QMC calculations, the lithium atom was systematically above the three adsorption sites. The energies were converged at each step to an accuracy of 1~$\times$~10$^{-6}$~Ry. To align the various curves, the non-interacting systems (i.e. graphite and lithium atom separately) were computed for each XC functional, and the binding energy was obtained according to Equation~\ref{eq:bindingenergy}.

\subsection{Molecular Density Functional Theory \label{sec:MDFT}}

Solvation free energies were computed with MDFT\cite{jeanmairet_molecular_2013,ding_efficient_2017,jeanmairet_study_2019}  while the polarisability of the graphite sheets was handled using fluctuating Gaussian charges method\cite{siepmann_influence_1995,scalfi_charge_2020,jeanmairet_study_2019}.
MDFT is a flavor of classical density functional theory (cDFT) developed to   study the solvation properties of molecular solutes into molecular solvents such as water or acetonitrile. The solvent is described by its density field $\rho({\bm r},\boldsymbol{\omega})$ which measures the average number per unit volume  of molecules with an orientation $\boldsymbol{\omega}$ at a given position ${\bm r}$. The solute acts a perturbation through an external potential $V_\text{ext}(\bm{r},\bm{\omega})$ causing the solvent to deviate from the homogeneous bulk fluid.

According to the cDFT principles\cite{evans_nature_1979,hansen_theory_2006}, there exists a unique functional, $F$, of the solvent density, $\rho$, that is equal to the solvation free energy at its minimum which is reached for the equilibrium solvent density. To find an expression for the functional, a common practice is to start by splitting it into the following sum 

\begin{align}
 F[\rho({\bm r},\boldsymbol{\omega})] & = F_\text{id}[\rho({\bm  r},\boldsymbol{\omega})]+F_\text{exc}[\rho({\bm r},\boldsymbol{\omega})] \nonumber \\ 
 & +\iint \rho({\bm r},\boldsymbol{\omega})V_\text{ext}({\bm r},\boldsymbol{\omega})d\bm{r}d\boldsymbol{\omega}. \label{eq:F=Fid+Fexc+Fext}
\end{align}

In Equation \ref{eq:F=Fid+Fexc+Fext}, the first term of the rhs is called ideal and corresponds to the entropic contribution of a non-interacting fluid with the same density. The second term is due to solvent-solvent interaction and is often called the excess term while the last term is due to solute-solvent interaction and thus called the external term.

Exact expressions exist for the ideal and external functionals that can be computed numerically. 
The excess part, however, requires approximations.  It can be expressed as an infinite Taylor expansion around the homogeneous bulk solvent density $\rho_b$,
\begin{equation}\label{excessF}
\begin{split}
F_\text{excess}[\rho] = 
& -\frac{k_BT}{2}\iiiint\Delta\rho({\bm r},\boldsymbol{\omega})c^{(2)}(\bm{r}-\bm{r}^\prime,\bm{\omega},\bm{\omega}^\prime) \\
& \Delta\rho({\bm r}^\prime,\boldsymbol{\omega}^\prime) d \bm{r} d\bm{\omega} d\bm{r}^\prime d\bm{\omega}^\prime + F_B[\rho].
\end{split}
\end{equation}
In Equation \ref{excessF}, $\Delta\rho(\bm r,\bm \omega) = \rho(\bm r,\bm \omega) - \rho_{b}$, $k_B$ is the Boltzmann constant, $T$ is the temperature and $c^{(2)}$ is the two-body direct correlation function of bulk solvent.
 $F_B$ is so-called bridge functional that collects all the terms higher than quadratic,  involving many-body direct correlation functions of the bulk solvent.
A common way to approximate the excess functional is to ignore the bridge functional, i.e. $F_B=0$, resulting to the ``HNC'' functional because it is equivalent to using the hypernetted chain (HNC) closure for the solute-solvent correlations in the molecular Ornstein-Zernike equation\cite{ding_efficient_2017}.
In this work, we use a very simple  bridge functional~\cite{borgis_simple_2020,borgis_accurate_2021} based on weighted density approximation that is known to correct well for the dramatic pressure overestimation of the HNC approximation.
Water is modeled with the SPC/E force field while the external potential $V_\text{ext}$ is created by the graphite electrodes and the lithium ion whose interactions are described as the sum of Lennard-Jones and electrostatic interactions. The Lennard-Jones and charge parameters of the 4 types of atoms involved are collected in Table \ref{tab:LJ-q-param}. 

\begin{table}
    \begin{center}
        \begin{tabular}{c|c|c|c}
            \hline
            Atoms & $\sigma \text{ (\AA)}$ & $\epsilon $ \text{(kJ/mol)} & charge (e) \\ \hline
             O &  3.166 & 0.65 & -0.8476  \\ \hline
            H & 0 & 0 & 0.4238 \\ \hline
            Li\cite{aqvist1990} &  2.216 &  0.07648 & 1 \\ \hline
            C\cite{werder2003} & 3.214 & 0.2364 & \text{Fluctuating} \\ \hline
        \end{tabular}
    \end{center}
            \caption{Force-field parameters used in the MDFT simulations. Mixed parameters are computed using the Lorentz-Berthelot rules (except for the C-Li interaction, which does not affect the MDFT results). } 
   \label{tab:LJ-q-param}            
    \end{table}

The electrode charges are represented with Gaussian charge distributions  with a width of 0.40~\AA.  These partial charges are calculated  by enforcing a uniform potential within the whole carbon electrode, with an overall electroneutrality constraint (hence forcing the total charge on the carbon to be equal to -1).\cite{siepmann_influence_1995}. Electrode charges are optimized self consistently with the functional minimization through an iterative scheme\cite{jeanmairet_study_2019}.

In the first step, the functional of Equation \ref{eq:F=Fid+Fexc+Fext} is minimized with no charges on the lithium and the carbon atoms. Then, carbon charges are optimized in the presence of the inhomogeneous water charge density and of the lithium cation.
The functional is minimized again but in the presence of lithium charge and the previously determined electrode charges.
The  process is repeated until it converges, with a convergence criterion of $5 \times 10^{-4}$ on the relative change in solvation free energy between two consecutive steps.

MDFT calculations were performed using an in-house  Fortran code and electrode charges are optimized using the constant potential molecular dynamics software MetalWalls\cite{marinlafleche_metalwalls_2020}.
We use a $24.672\times21.366\times40$ \AA$^3$ simulation box (the unit cell used in QMC is replicated twice in $x$ and $y$ directions) with $74\times64\times120$ grid nodes and an angular grid of 196 orientations per grid node. We run calculations for a distance $z$ between the electrode plane and the lithium varying between $z = 1.0$~\AA\ and $z = 10$~\AA\ with an increment of 0.2~\r{A} between 1~\r{A} and 6~\r{A} and of 0.5~\r{A} otherwise.

\begin{figure*}[ht!]
    \centering
    \includegraphics[width=\textwidth ]{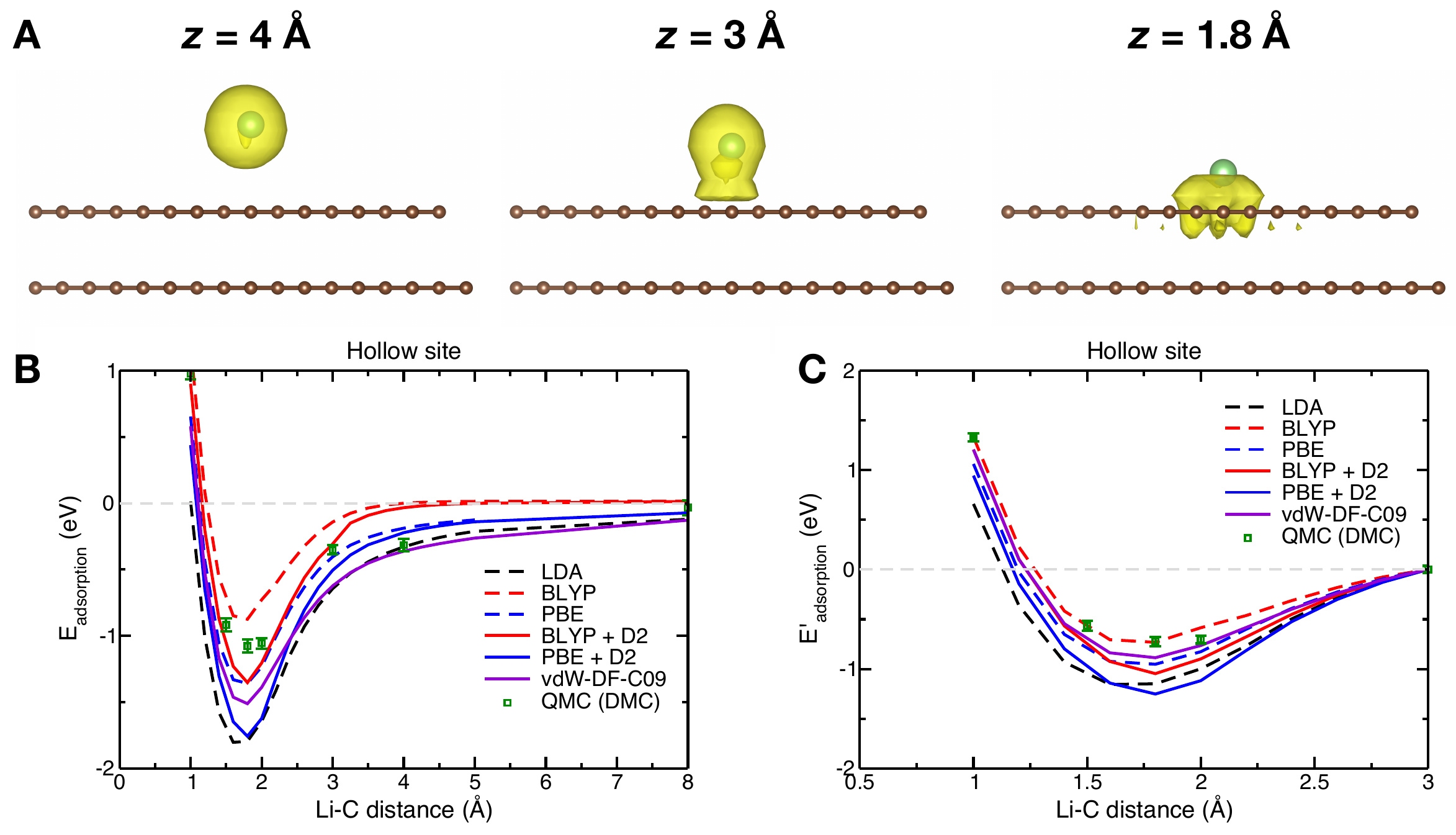}
    \caption{ A) Electron density around a Li atom adsorbed on a graphite substrate, computed via quantum Monte Carlo, for different atom--substrate distances. The densities were obtained by computing the overall electronic density of a system with a Li atom adsorbed on graphite, and subtracting the density of the isolated substrate, in absence of the adsorbed atom. All shown isosurfaces correspond to a density of 6 $\cdot 10^{-4}$ electrons/\AA{}$^3$. B) Comparison of the adsorption energies obtained with various DFT functionals and DMC for the adsorption of the lithium on the hollow site of graphite. C) Same as B) but substracting the adsorption energy at a distance of 3~\AA.}
    \label{fig:QMC}
\end{figure*}


\section{Results and discussion}

\subsection{Benchmark of the DFT functionals with QMC}

\begin{figure*}
    \centering
    \includegraphics[width=\textwidth ]{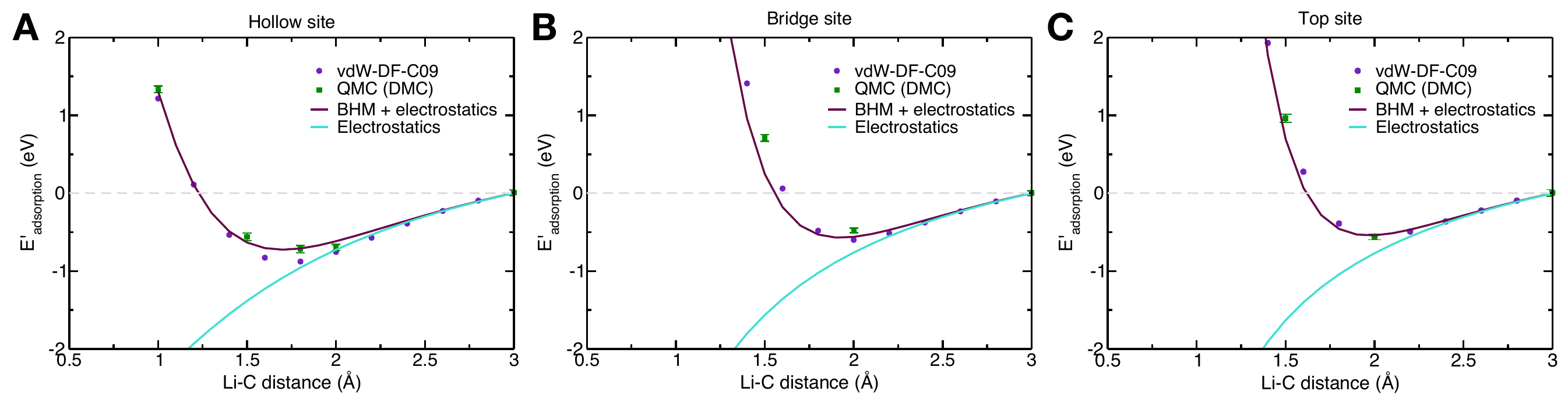}
    \caption{ Comparison of the fitted potential with the QMC and vdW-DF-C09 energies.}
    \label{fig:fitting}
\end{figure*}

We first discuss the QMC results for the three adsorption sites. QMC total energies and adsorption energies are reported in Table \ref{tab:qmc_abs}. 

\begin{table}[]
    \centering
    \begin{tabular}{c|c|c|c}
    \hline
                  & $E_{VMC}$ (Ha) & $E_{DMC}$ (Ha) & $E_{ads}$ (eV) \\
    \hline
    Graphite     & -567.2021(4)   &   -568.531(1)  &  --             \\
    Li atom       & -0.198050(9)   &   -0.198314(3)   &  --             \\
    \hline
    Hollow site                 \\  
    \hline
    1.0           & -567.3573(5)   &   -568.693(1)  &   0.98(4)       \\
    1.5           & -567.4297(4)   &   -568.763(1)  &  -0.92(5)       \\
    1.8           & -567.4341(4)   &   -568.769(1)  &  -1.08(5)       \\
    2.0           & -567.4313(5)   &   -568.768(1)  &  -1.06(4)       \\
    3.0           & -567.4006(5)   &   -568.742(1)  &  -0.35(4)       \\
    4.0           & -567.4011(4)   &   -568.741(1)  &  -0.32(5)       \\
    8.0           & -567.3935(4)   &   -568.730(2)  &  -0.03(6)       \\
    \hline
    Top site \\
    \hline
    1.5           & -567.3663(4)   &   -568.705(2)  &   0.65(5)       \\
    2.0           & -567.4226(4)   &   -568.761(1)  &  -0.87(4)       \\
    3.0           & -567.4005(5)   &   -568.740(2)  &  -0.31(4)       \\
    \hline
    Bridge site \\
    \hline
    1.5           & -567.3799(4)   &   -568.716(2)  &   0.36(4)       \\
    2.0           & -567.4251(4)   &   -568.760(1)  &  -0.83(3)       \\
    3.0           & -567.3999(4)   &   -568.742(1)  &  -0.35(3)       \\
    \hline
    \end{tabular}
    \caption{QMC energies for the adsorption of lithium on graphite, for the three different sites for several lithium-carbon distances  reported in the first  column (in \AA). Second column are VMC results while third column are DMC results. The last column are the  adsorption energies computing with DMC using equation~\ref{eq:bindingenergy}.}
    \label{tab:qmc_abs}
\end{table}


From Table \ref{tab:qmc_abs}, it appears clearly that the adsorption energies are larger for the top and bridge sites than for the hollow site for distances of 1.5 and 2.0~\AA\, in good agreement with previous eDFT results from the literature and from the current study. At a larger distance of 3.0~\AA\, the three sites display similar energies, which shows that the difference between them has a short-range character. Due to the high computational cost of DMC, further lithium-carbon distances were only considered for the hollow site. We obtained a binding energy $E_b$ of -1.08~eV for a lithium-surface distance of 1.8~\AA. We also observe a somewhat peculiar behavior since the adsorption energy is rather similar for distances of 3 and 4~\AA. By analyzing the corresponding electronic densities as shown on Figure \ref{fig:QMC}A, we observe that this correspond to the region in which the electron transfer occurs. For distances lower than 3~\AA\, the energies correspond to the adsorption of a lithium ion on a polarized carbon surface while for distances greater than 4~\AA\, the system corresponds to neutral lithium atom and carbon material. 


The adsorption profiles obtained for the hollow site for the LDA, BLYP, PBE, BLYP+D2, PBE+D2 and dW-DF-C09 functionals are compared with the DMC benchmark on Figure \ref{fig:QMC}B. The LDA results in a strong overbinding, which is expected. The comparison with the other functionals is more surprising. The QMC results lie between the BLYP and the BLYP+D2, while all the other functionals predict too low adsorption energies. However, as noted by Valencia {\it et al.}, the binding energy (defined as the minimum of the adsorption energy profile) should be approximately given by~\cite{valencia2006a}
\begin{equation}
    E_b \approx E_b({\rm Li^+})-(IP[{\rm Li}]-W_f[{\rm Graphite}])
\end{equation}
\noindent where $E_b$(Li$^+$) is the binding energy of the lithium {\it ion}, $IP$ its ionization potential and $W_f$ the work function of graphite. The observed variation may be linked to compensation errors between the various terms.


In the present work, we are mostly interested in the adsorption of the lithium ion. Consequently, we performed a second comparison of the various functionals in which the energy at $z$~=~3~\AA\ is substracted. The results are shown on Figure~\ref{fig:QMC}C. The discrepancy between the various functionals is somewhat lower. The best agreement is now obtained with BLYP, PBE and the vdW-DF-C09 functionals, while the others predicts overbinding. This points towards an overestimation of the dispersion effects when using the D2 correction. Indeed, in the case of the lithium ion, only two semi-core electrons take part in the interaction, which should result in a very weak dispersion term. The  vdW-DF-C09 functional, which accounts for these effect explicitly and not through a parameterized term, seems to better catch the interactions. 

\subsection{Fitting the carbon--lithium potential}

In order to incorporate solvent effects, it is necessary to develop accurate classical interaction potentials. It is not possible to fit them directly on the QMC calculations due to the limited number of data. Instead, we pick the most accurate functional, vdW-DF-C09, and calculate the Li--graphite binding energy for a large number of distances. The intermolecular interaction should in principle account for four different effects: electrostatics, polarization, short-range repulsion and dispersion. In our electrostatic model, the two former effects are explicitly introduced through the use of a +1 point charge on the lithium and of the calculation of partial (Gaussian) charges on the carbon atoms. These partial charges are calculated for each lithium-carbon distance using the same methodology as described in \ref{sec:MDFT}.

Concerning the short-range repulsion and the dispersion effects, the two main potentials used in the literature are the Lennard-Jones and the Born-Huggins-Mayer (BHM) ones. However, it appears that the electrostatic interaction was sufficient to account for the attractive part of the binding energy, as shown on Figure \ref{fig:fitting}. The fitted potential should therefore add very few, if no contribution for the dispersion interaction, which agrees with the previous observation on the use of dispersion-corrected functionals. A well-known drawback of the Lennard-Jones potential is that it is not possible to fit the short-range repulsion and the dispersion term separately since they both involve the same parameters. We have therefore chosen a BHM potential instead, which analytical form is:
\begin{equation}
    V_{\rm BHM}(r) = A \exp{\left(-b r\right)}-\frac{C_6}{r^6}
\end{equation}
\noindent where $A$, $b$ are the parameters describing the intensity and the range of the repulsion interaction, while $C_6$ is the dipole-dipole dispersion interactions. In principle higher order terms could be included for dispersion, but as discussed above this term is almost negligible in the case of the lithium ion. The fitted potential reproduces with a very high accuracy the vdW-DF-C09 as well as the QMC results (on which it was not fitted) for the three types of adsorption sites as shown on Figure \ref{fig:fitting}. The corresponding parameters are $A$~=~91.17, $b$~=~2.518 and $C_6$~=~1.107 (all numbers are given in atomic units). 

\subsection{Adsorption of the lithium ion in the presence of water}

\begin{figure}
    \centering
    \includegraphics[width=\columnwidth ]{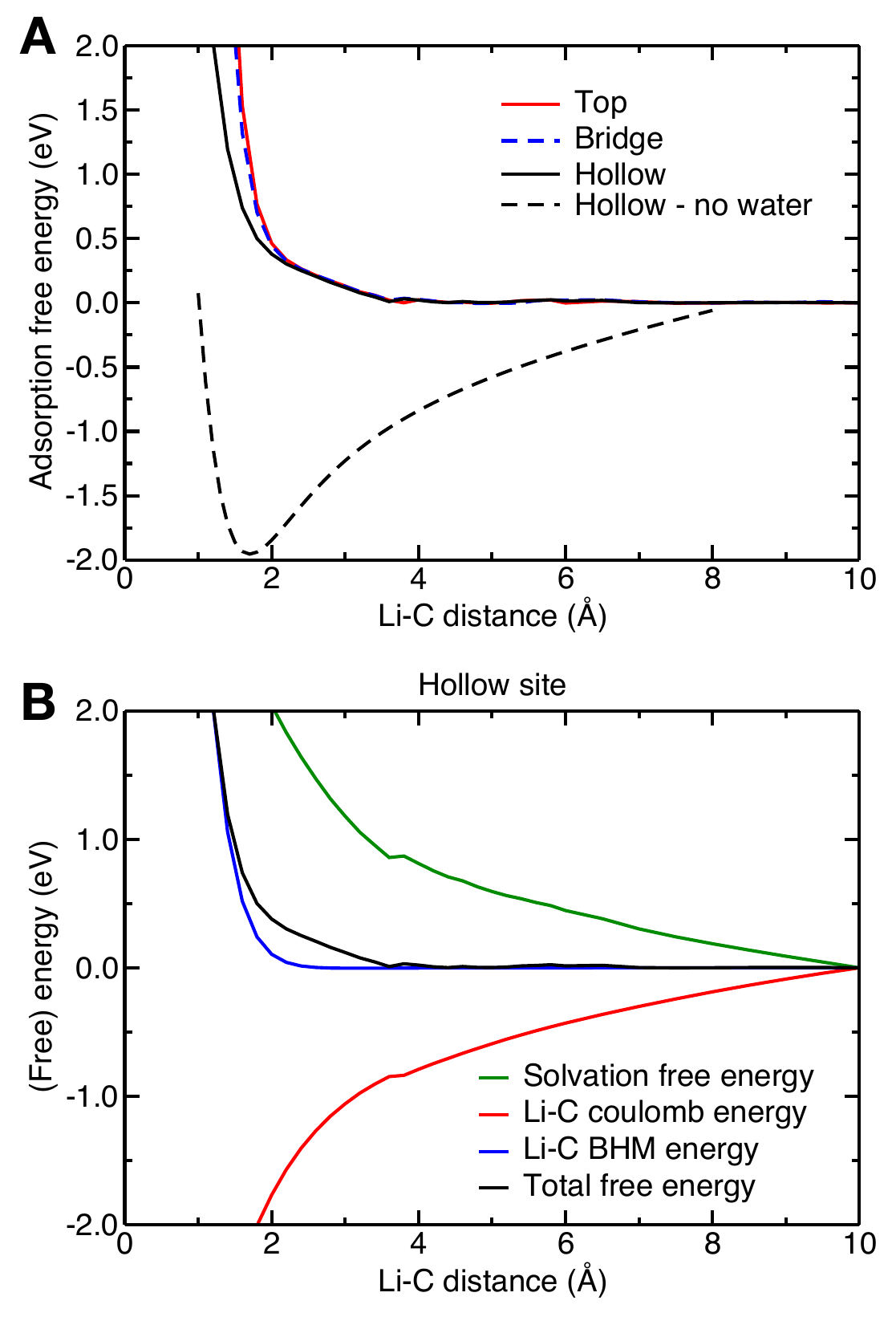}
    \caption{ A) Adsorption free energy for a lithium ion on the graphite surface in the presence of water, computed using MDFT, for the three adsorption sites. The energy variation in the absence of water is also shown for comparison. B) Contributions to the total free energy for the hollow adsorption site in the MDFT calculation.}
    \label{fig:MDFTenergy}
\end{figure}

The fitted potential can directly be used in any classical molecular simulation, such as MD. Since we focus here on the adsorption free energy  of the lithium ion on the carbon surface we prefer to use MDFT which is a computationally  more efficient alternative. The solvation free energy of a single system can be computed within a few minutes on a single CPU while  it would require tens of CPU hours with MD. The free energy profile obtained for the three adsorption sites in the presence of liquid water are shown on Figure \ref{fig:MDFTenergy}A. The profiles are very different from the gas phase results. The minimum at $\approx$~1.8~\AA\ completely disappears and is replaced by a strongly repulsive wall. This shows that there is no preferential adsorption of lithium on the graphite surface in aqueous phase.  

This result is in qualitative agreement with a recent MD study on the adsorption of hydrated ions on graphene,~\cite{loche2018a} which provided a repulsive free energy profile over the whole range of considered distances. Yet, the latter study did not include any Coulombic interaction between the ion and the carbon surface, which is the main driving force for adsorption in the gas phase as discussed above. It is thus interesting to examine the various contributions to the total free energy, which are provided on Figure \ref{fig:MDFTenergy}B. We observe that the electrostatic attraction between Li and C is in fact counterbalanced by the solvation free energy. The latter contains the electrostatic interactions of the water molecules with both the lithium and the graphite surface, which results in strong screening effect. The extent of this screening effect was studied in details in a recent study focused on gold surfaces:~\cite{pireddu2021a} The presence of the water molecules strongly impacts the polarization of the surface. Consequently, the total free energy is almost equal to the BHM contribution over the whole range of distances, except between 2 and 5~\AA\ where the solvation free energy overcomes the ion-surface Coulombic interaction, resulting in a more repulsive potential.

\begin{figure*}
    \centering
    \includegraphics[width=\textwidth ]{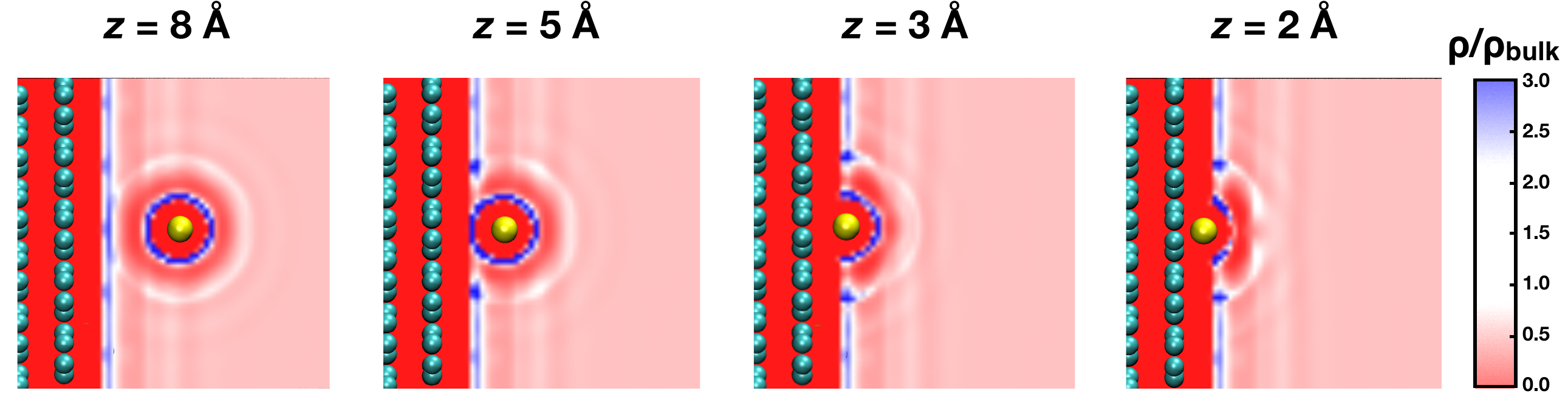}
    \caption{ Projection of the solvent densities computed using MDFT for various lithium--carbon distances.}
    \label{fig:MDFdensities}
\end{figure*}

The effect of the solvent can be further analyzed by plotting the density profiles for various distances between the ion and the surface (Figure \ref{fig:MDFdensities}). At $z$~=~8~\AA, two regions with larger densities emerge, corresponding to the surface adsorbed water molecules at a distance of 3~\AA\ from the surface~\cite{jeanmairet_study_2019} on the one hand and to the lithium ion first solvation shell on the other hand. At a distance of 5~\AA\, the solvation shell starts to overlap with the adsorbed layer at close contact to the electrode, which results in small depletion zones in the latter. These depletion zones remain observable at smaller distances, but the impact on the free energy becomes negligible w.r.t the large short-range repulsion between the carbon and the lithium. 

\section{\label{sec:Concl}Conclusion}

Two main results were obtained in this work. Firstly, we provide an accurate set of benchmark energies for the adsorption of lithium on a graphite surface, using an accurate QMC approach. This data can be used in future works to test the accuracy of XC functionals for such important problems in surface science, or to parameterize new force fields between the two species. Here we obtained a very good agreement using a BHM potential, but different approaches may be proposed in the future. Secondly, we used the parameterized force field in order to compute the free energy profile for the adsorption of a lithium ion at the graphite surface in the presence of water as a solvent. MDFT was preferred over classical MD for this calculation due to its much lower computational cost. We showed that the low energy minimum obtained in the gas phase completely vanishes, resulting in an overall repulsive profile over the whole range of studied lithium--carbon distances. This is due to the screening of the attractive Coulombic term by the water molecules on the one hand, and on the other hand to the interferences between the densities corresponding to the first adsorbed layer on carbon and to the first solvation shell of the cation. 

The proposed multi-scale approach may be extended to a large variety of systems in the future. In particular, it would be interesting to study how adsorption varies within the alkali and alkali-earth cationic series. The study of larger organic ions would be very useful for the scope supercapacitors devices, but this would require additional work to account accurately for the flexibility of the ion. Finally, the   study of more polarizable electrode materials such as gold or platinum would be relevant in the context of catalysis. Changing metal would lead to very different adsorption properties for the solvent, so that one can expect large differences in the adsorption free-energy profiles of ions at electrodes.

\begin{acknowledgments}

This project has received funding from the European Research Council (ERC) under the European Union's Horizon 2020 research and innovation programme (grant agreement No. 771294). This work was granted access to the HPC resources of CINES under Allocation A0100910463 made by GENCI and benefited from the `Grand Challenge Jean Zay' program. We acknowledge support from EoCoE, a project funded by the European Union Contracts No. H2020-EINFRA-2015-1-676629 and H2020-INFRAEDI-2018-824158.

\end{acknowledgments}

\section*{Data Availability}
The input files used in this study for the QMC, DFT and classical simulations are openly available in the repository https://gitlab.com/ampere2, together with the reference energies for the QMC calculations.

\bibliography{Li_graphite}

\end{document}